\documentclass[aps,prb,reprint,showpacs,showkeys]{revtex4-2}
\usepackage{graphicx} % Include figure files
\usepackage{amsmath} % Useful mathematical tools
\usepackage{mathtools}
\usepackage{amssymb}
\newcommand*{\permcomb}[4][0mu]{{{}^{#3}\mkern#1#2_{#4}}}

\newcommand*{\comb}[1][-1mu]{\permcomb[#1]{C}}
\usepackage[version=4]{mhchem} % for chemical equations
\usepackage{bm} % bold mathi
\usepackage{natbib}
\usepackage{natmove}
\usepackage[colorlinks=true, linkcolor=blue, citecolor=blue, urlcolor=blue, linktoc=page, bookmarks=false, pdfstartview={FitH}, pdfborder={0 0 0.0 [3 3]}]{hyperref} % HyperLink
\usepackage{color} % For colored text
\usepackage{dcolumn} % Align table columns on decimal point
\newcolumntype{.}{D{.}{.}{2.5}}
\newcolumntype{-}{D{.}{.}{4.0}}
\usepackage{makecell}
\usepackage{multirow}
\usepackage{cleveref}
\usepackage{easyReview} % Review mode
\crefname{figure}{Fig.}{Figs}
\crefname{table}{Table}{Tables}
\renewcommand{\today}{\number\day \space \ifcase \month \or January\or February\or March\or April\or May\or June\or July\or August\or September\or October\or November\or December\fi \space \number\year} % Date
\def\m1r{\multicolumn{1}{r}}
\newcolumntype{P}[1]{>{\centering\arraybackslash}p{#1}}
\begin{document}
% ========== TITLE ==========
\title{Unsupervised Deep Neural Network Approach To Solve Bosonic Systems}
%Magnetic and Topological Properties of Electron Doped SrIrO$_3|$SrTiO$_3$ Interface}
% ====================
% ========== AUTHORS AND AFFILIATIONS ==========
\author{Avishek \surname{Singh}}
\email[Email: ]{avishek@iiserb.ac.in}
% --------------------
\author{Nirmal \surname{Ganguli}}
\email[Email: ]{NGanguli@iiserb.ac.in}
\affiliation{Department of Physics, Indian Institute of Science Education and Research Bhopal, Bhauri, Bhopal 462066, India}
% ====================
\date{\today}
% ========== ABSTRACT ==========
\begin{abstract}
The simulation of quantum many-body systems poses a significant challenge in physics due to the exponential scaling of Hilbert space with the number of particles. Traditional methods often struggle with large system sizes and frustrated lattices. In this research article, we present a novel algorithm that leverages the power of deep neural networks combined with Markov Chain Monte Carlo simulation to address these limitations. Our method introduces a neural network architecture specifically designed to represent bosonic quantum states on a 1D lattice chain. We successfully achieve the ground state of the Bose-Hubbard model, demonstrating the superiority of the adaptive momentum optimizer for convergence speed and stability. Notably, our approach offers flexibility in simulating various lattice geometries and potentially larger system sizes, making it a valuable tool for exploring complex quantum phenomena. This work represents a substantial advancement in the field of quantum simulation, opening new possibilities for investigating previously challenging systems.
\end{abstract}
% ====================
 % Physics and Astronomy Classification Scheme
\maketitle
% ====================
% ========== INTRODUCTION ==========
\section{\label{sec:intro}Introduction}
The concept of the wave function is a crucial aspect of quantum mechanics but can be challenging to understand in classical terms. The wave function represents a comprehensive mathematical object that encodes all the information of a quantum state, whether a single particle or a complex molecule. Theoretically, an exponential amount of data is required to fully describe a generic many-body quantum state. However, many physical many-body systems can be characterized by an amount of information that is much less than the maximum capacity of the corresponding Hilbert space. This is due to a limited amount of quantum entanglement and a small number of physical states within these systems. This allows modern methods to solve the many-body Schrödinger’s equation with fewer classical resources.

The possible states of a many-body system are described by a Hilbert space whose dimension grows exponentially with the number of particles and one-body states. This presents a significant challenge for physicists studying many-body problems as it becomes increasingly difficult to simulate these systems accurately. Consequently, physicists resort to using approximate techniques to carry out simulations (e.g., dynamical mean-ﬁeld theory \cite{antoine1996}, density matrix renormalization group (DMRG)\cite{Schollwock2005}). The ground state of a many-body system is a quantum object that has been extensively studied for various reasons. The particles in the system typically occupy the lowest energy states according to the Aufbau principle. The ground state notably provides information about observable changes demonstrating quantum phase transitions.

Recent developments in artificial neural networks have revolutionized the field of artificial intelligence. Coupled with high-performance computers, ANNs have been able to perform complex tasks in various fields, including engineering and science. In particular, machine learning techniques, such as pattern recognition, have found multiple applications in physics research. ANNs have proven to be an effective tool for handling complex data sets, and their use in the study of many-body systems has opened up new avenues for research. Trained ANNs can discriminate different phases of numerically or analytically obtained many-body states\cite{tomoki2016, Carrasquilla2017, VanNieuwenburg2017, Zhang2017, Tomi2017, Broecker2017, Tanaka2017, Broecker2017-1, Kelvin2018, Zhang2018, Mano2017}.

A groundbreaking study demonstrated the potential of Artificial Neural Networks (ANNs) to solve quantum many-body problems \cite{Carleo2017}. Using a restricted Boltzmann machine, researchers accurately represented the ground states and time evolutions of quantum Ising and Heisenberg models. The ANN fed a spin configuration optimized to produce wave functions minimizing energy. This work highlights how machine learning can address the challenge of describing the exponentially complex correlations within many-body wave functions. The variational representation of quantum states using ANNs enabled studying complex interacting systems' ground states and unitary time evolution. Due to the exponential scaling of Hilbert space with particle count, physicists often rely on approximations like Monte Carlo \cite{Sandvik1999} or tensor network methods \cite{Orus2019}. ANNs offer a promising alternative by reducing the computational cost of simulating many-body systems. Their ability to identify patterns in data has also proven helpful in detecting phase transitions in systems like magnetic materials and superconductors. This demonstrates the significant promise ANNs hold for tackling previously intractable physics problems.

Motivated by Ref. \cite{Hangleiter2020, Carleo2017}, in this work, we present an algorithm developed using a deep neural network and Markov chain Monte Carlo, leveraging variational principle, a fundamental principle of quantum mechanics. We show that a fully connected feed-forward neural network could learn the complex many-body quantum system parameters and construct the entire quantum state. We demonstrated that the fully connected feed-forward neural network with one input, one hidden, and one output layer could predict the ground state of the Bose-Hubbard model on a one-dimensional lattice chain. Regarding the optimization of the neural network, we have formulated three optimization algorithms: steepest descent, adaptive gradient, and adaptive momentum. We demonstrated that the adaptive gradient and the adaptive momentum optimization algorithms help in faster optimization of the neural network and help in overcoming the limitations of the steepest descent algorithm to optimize the neural network parameters. Also, the presented algorithm is robust and capable of simulating any shape or size of the lattice, which means it can also simulate the frustrated lattices, for which most traditional methods like quantum Monte Carlo fail.
% ====================

% ========== Algorithm ======
\section{\label{sec:Algorithm}Algorithm}
The general schematics of the developed algorithm consist of three parts: (a) Quantum System, (b) Sampling Algorithm and (c) Predictive Algorithm.

%----- Quantum System ------------------------------------------------------------
\subsection{\label{subsec:Algorithm-Quantum_System}The Quantum System and Lattice}

\begin{figure}
    \centering
    \includegraphics[scale = 0.24]{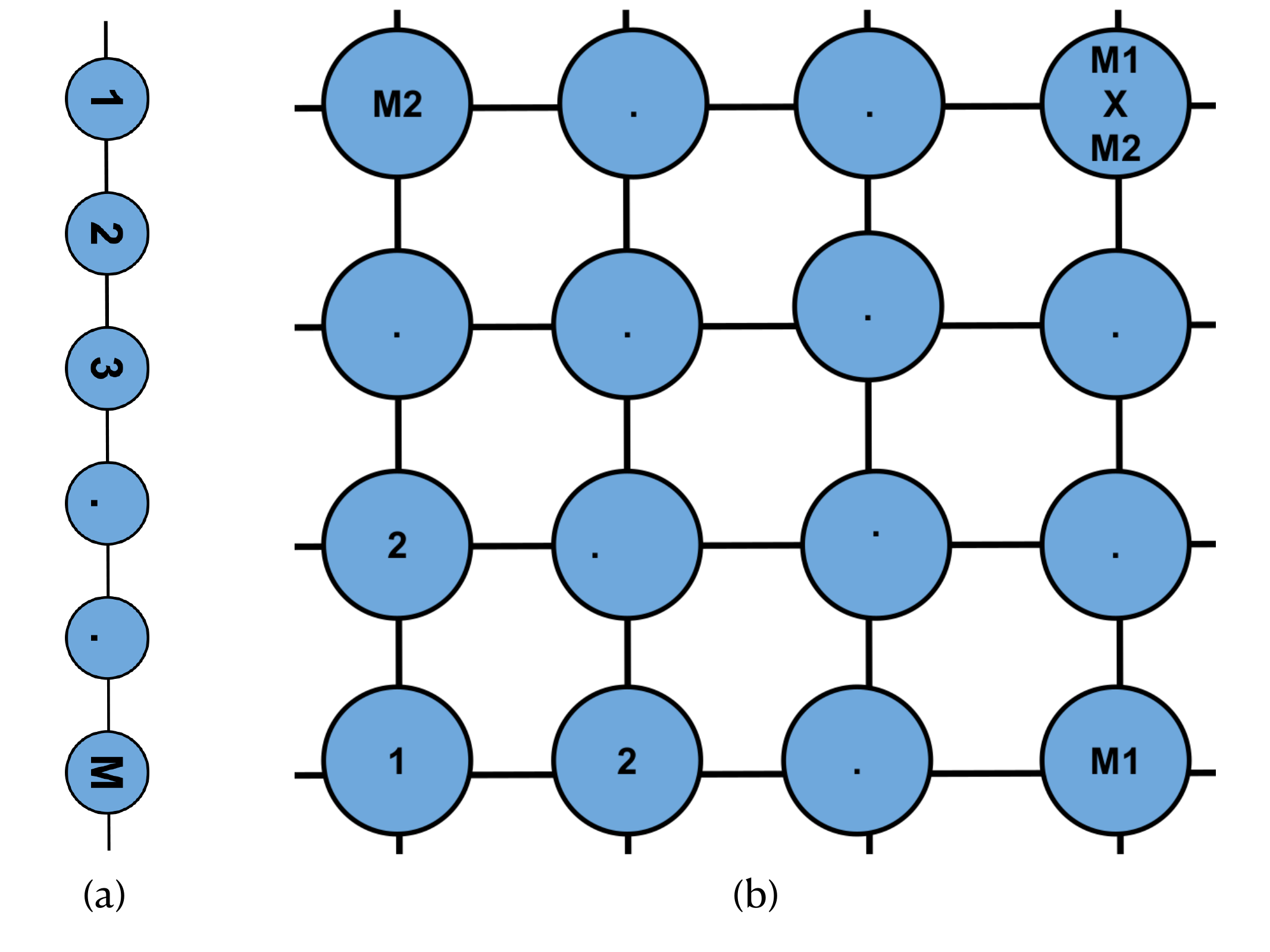}
    \caption{1D and 2D lattices to describe many-body formalism for a quantum system.}
    \label{fig:Algorithm-Quantum_System-Lattice}
\end{figure}
Before defining the many-body formalism for quantum systems, we will first define the lattice on which the formalism will be constructed. The \cref{fig:Algorithm-Quantum_System-Lattice}(a) represents the 1D lattices with $M$ number of sites while \cref{fig:Algorithm-Quantum_System-Lattice}(b) represents 2D lattice with $M_1$ sites along the x-axis and $M_2$ sites along the y-axis, which makes the total number of sites in the 2D lattice $M = M_1 \times M_2$. In general, we label each site of the lattice starting from $1$ to $M$ with the total number of sites $M$ irrespective of lattice type or dimension as represented in \cref{fig:Algorithm-Quantum_System-Lattice}. This means that regardless of type or dimension, our lattice will be represented as a linear chain of sites to the deep neural network. This modification has been made to use the layered structure of the deep neural network, represented in \cref{fig:Algorithm-Predictive-DeepNeuralNetwork}, as we will input the particle occupation per site as a vector to the input layer of the neural network.

For a given lattice with $M$ sites occupied by $N$ bosons, a quantum many-body state of particles on a lattice can be expanded by Fock states as,
\begin{align}
    \label{eq:Algorithm-Quantum_System-FockState}
    | \Psi \rangle &= \sum_n \phi(n_1, n_2, n_3, .... , n_M) | n_1, n_2, n_3, .... , n_M \rangle \\
                   &= \sum_n \phi(\bm{n}) | \bm{n} \rangle \nonumber
\end{align}
where $\bm{n} = (n_1, n_2, n_3, ...., n_M)$ represents the distribution of particles on the lattice sites, and $M$ is the number of sites. That means $n_1$ is the number of particles on lattice site-1, $n_2$ is the number of particles on lattice site-2 and so on. $\phi(\bm{n})$ is in general a complex number.

In \cref{eq:Algorithm-Quantum_System-FockState}, the occupation numbers $\bm{n}$ of any quantum state are generated by sampling technique as discussed in \cref{subsec:Algorithm-Sampling}, and $\phi(\bm{n})$ is predicted by predictive algorithm given in \cref{subsec:Algorithm-Predictive}. Once $\phi(\bm{n})$ is predicted using a deep neural network, a full many-body quantum state can be constructed for a given occupation state. The total number of possible Fock state ($N_\text{Tot}$) are calculated using permutation and combination theory as,
\begin{align}
    \label{eq:Algorithm-Quantum_System-Total_Fock_States}
    N_\text{Tot} &= \comb{(N + M -1)}{N} = \binom{N+M-1}{N} \\
                 &= \frac{(N + M -1)!}{N! (M-1)!} \nonumber
\end{align}
The total number of particles $N = \sum_{i=1}^M n_i$ must satisfy this condition.

For a given Hamiltonian ($H$) and many-body quantum state $|\Psi\rangle$, the total energy of the quantum system is given by the expectation value of Hamiltonian as,
\begin{align}
    \label{eq:Algorithm-Quantum_System-ExpectationValue}
    \langle \hat{H} \rangle &= \frac{\langle \Psi | \hat{H} | \Psi \rangle} {\langle \Psi | \Psi \rangle} \nonumber \\
                            &= \frac{\sum_{\bm{n}, \bm{n'}} \phi^*(\bm{n}) \langle \bm{n} | \hat{H} | \bm{n'} \rangle \phi(\bm{n'})}{\sum_{\bm{n}} |\phi(\bm{n})|^2}
\end{align}
where $| \Psi \rangle$ is the many-body quantum state, $\phi(\bm{n})$ is a complex coefficient, and $|\bm{n}\rangle$ is the occupation state. 
%---------------------------------------------------------------------------------

%----- Sampling ------------------------------------------------------------------
\subsection{\label{subsec:Algorithm-Sampling} The Sampling Algorithm: Markov Chain Monte Carlo}

The occupation states $|\bm{n}\rangle$ are sampled from Hilbert space using importance sampling with two methods: \textit{Local Sampler}, and second is \textit{Exchange Sampler}. To begin with, let's take an initial randomly generated quantum state $| \Psi \rangle = \sum_n \phi(n_1, n_2, n_3, .... , n_M) | n_1, n_2, n_3, .... , n_M \rangle$ for which occupation state $ |\bm{n}\rangle$ is given as,
\begin{equation}
    \label{eq:Algorithm-Sampling-OcuupationState}
    |\bm{n}\rangle = | n_1, n_2, n_3, ..., n_i, ..., n_j, ..., n_M \rangle
\end{equation}

The \textit{Local Sampler} works on one local degree of freedom, which means it will change the particle occupation only at one site of the lattice to construct the new occupation state. The rules are as follows, 
\begin{enumerate}
    \item This sampler acts locally only on one local degree of freedom  $n_i$ and proposes a new state as follows,
            \begin{equation}
                \label{eq:Algorithm-Sampling-OcuupationState_local}
                |\bm{n'}\rangle = | n_1, n_2, n_3, .., n'_i, ..., n_M \rangle
            \end{equation}
            where $n'_i \neq n_i$  
    \item The transition probability associated with this sampler can be decomposed into two steps. First is one of the site indices $i \in \{1, ......., M\}$ is chosen with uniform probability, and the second is among all the possible $n'_i$ that can take, one of them is chosen with uniform probability.
    \item In the case of bosons, We randomly create or annihilate a particle on a chosen lattice site with uniform probability. Also, whether to create or annihilate the particle on the chosen site is decided by uniform probability.
\end{enumerate}
The \textit{Exchange Sampler} works on two local degrees of freedom, which means it will change particle occupation on two lattice sites to construct the new occupation state. The rules are as follows,
\begin{enumerate}
    \item This sampler acts locally on two degrees of freedom $n_i$ and $n_j$ and proposes a new state as follows,
            \begin{equation}
                \label{eq:Algorithm-Sampling-OcuupationState_Exchange}
                |\bm{n'}\rangle = | n_1, n_2, n_3, ..., n'_i, ..., n'_j, ..., n_M \rangle
            \end{equation}
            where $n'_i \neq n_i$ and $n'_j \neq n_j$
    \item The transition probability associated with this sampler can be decomposed into two steps. First is a pair of indices $(i, j) \in \{1, ......., M\}$, such that $\text{dist}(i,j \leq d_\text{max})$ is chosen with uniform probability and the second is among all the possible $n'_i$, $n'_j$ can take, one of them is chosen with uniform probability.
    \item In the case of bosons, We choose two sites $(i, j) \in \{1, ......., M\}$ with uniform probability and at site $i$, we create a particle, and at site $j$, we annihilate a particle.
\end{enumerate}

Using \textit{Local Sampler} method, the total number of bosons can not be fixed while \textit{Exchange Sampler}, the total number of bosons can be fixed throughout the calculation.

The probability ($P_{(\bm{n} \to \bm{n'})}$) of accepting the newly proposed state$|\bm{n'}\rangle$ from initial state $|\bm{n}\rangle$ using sampler is give as,
\begin{equation}
    \label{eq:Algorithm-Sampling-AcceptanceProb}
    P_{(\bm{n} \to \bm{n'})} = \text{min} \left[ 1, \left| \frac{\phi(\bm{n'})}{\phi(\bm{n})} \right|^2 \right]
\end{equation}
Here, note that the squared probability $\left| \frac{\phi(\bm{n'})}{\phi(\bm{n})} \right|^2$ allows us to simulate even frustrated lattices by avoiding sign problem. With this probability, the sampling probability distribution of occupation state $|\bm{n}\rangle$ becomes as,
\begin{equation}
    \label{eq:Algorithm-Sampling-ProbDist}
    p(\bm{n}) = \frac{\left| \phi(\bm{n}) \right|^2}{\sum_{\bm{n'}} \left| \phi(\bm{n'}) \right|^2}
\end{equation}

The expectation value of Hamiltonian ($\langle \hat{H} \rangle$) given by \cref{eq:Algorithm-Quantum_System-ExpectationValue} using probability distribution given in \cref{eq:Algorithm-Sampling-ProbDist} can be modified as,

\begin{align}
    \label{eq:Algorithm-Quantum_System-ExpValModified}
    \langle \hat{H} \rangle &= \frac{\sum_{\bm{n}, \bm{n'}} \phi^*(\bm{n}) \langle \bm{n} | \hat{H} | \bm{n'} \rangle \phi(\bm{n'})}{\sum_{\bm{n}} |\phi(\bm{n})|^2} \nonumber \\
                            &= \sum_{\bm{n}, \bm{n'}} \frac{\left| \phi(\bm{n}) \right|^2}{\sum_{\bm{n'}} \left| \phi(\bm{n'}) \right|} \langle \bm{n} | \hat{H} | \bm{n'} \rangle \frac{\phi(\bm{n'})}{\phi(\bm{n})} \nonumber \\
                            &= \sum_{\bm{n}, \bm{n'}} p(\bm{n}) \langle \bm{n} | \hat{H} | \bm{n'} \rangle \frac{\phi(\bm{n'})}{\phi(\bm{n})} \\
                            &\simeq \left< \sum_{\bm{n'}} \langle \bm{n} | \hat{H} | \bm{n'} \rangle \frac{\phi(\bm{n'})}{\phi(\bm{n})} \right>_{N_T} \nonumber \\
                            &\simeq \langle \Tilde{H} \rangle_{N_T} \nonumber
\end{align}

This modified expectation value of Hamiltonian ($\langle \hat{H} \rangle$) will serve as the total energy of the system and the cost function of the deep neural network.
%---------------------------------------------------------------------------------

%----- Predictive Model ----------------------------------------------------------
\subsection{\label{subsec:Algorithm-Predictive} The Predictive Algorithm: Deep Neural Networks}

\begin{figure}
    \centering
    \includegraphics[scale = 0.45]{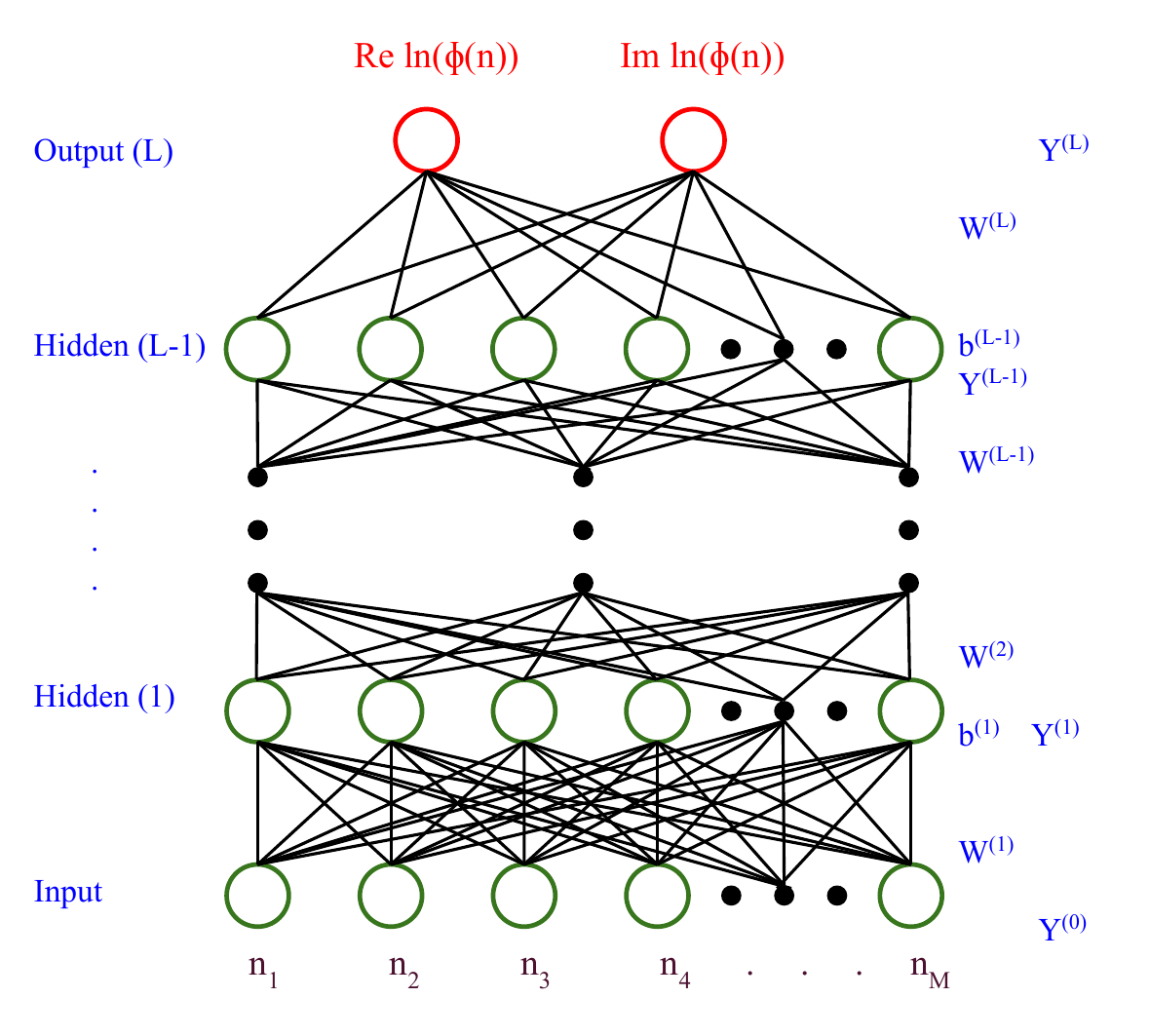}
    \caption{A general Schematic diagram of the architecture of the fully connected feed-forward deep neural network. The input and output of this neural network are represented as $Y^{(0)}$ and $Y^{(L)}$, respectively.}
    \label{fig:Algorithm-Predictive-DeepNeuralNetwork}
\end{figure}

\cref{fig:Algorithm-Predictive-DeepNeuralNetwork} depicts the architecture of a fully connected feed-forward deep neural network. The neural network consists of input, hidden, and output layers, and each layer consists of neurons or nodes represented by circles in the schematic. In \cref{fig:Algorithm-Predictive-DeepNeuralNetwork}, the neurons of input and output layers are denoted by $Y^{(0)}$ and $Y^{(L)}$, respectively, and the units of hidden layers are denoted by $Y^{(1)}$, $Y^{(2)}$, ....., $Y^{(L-1)}$. These $Y^{(0)}$, $Y^{(1)}$, ....., $Y^{(L)}$ also represent the output of each neuron in the layer. The number of neurons in the input and output layer depends on the problem. For example, let us say we want to construct the quantum state $|\Psi(\bm{n}) \rangle = \sum_n \phi(\bm{n}) | \bm{n} \rangle$. To construct this quantum state, we will predict $\phi(\bm{n})$ from the neural network and occupation state $| \bm{n} \rangle$ will be input for the neural network. As $\phi(\bm{n})$ is a complex number, it will consist of real and imaginary parts; thus, in  \cref{fig:Algorithm-Predictive-DeepNeuralNetwork} output layer has two neurons. From \cref{subsec:Algorithm-Quantum_System}, we know that occupation states $| \bm{n} \rangle$ will have values equal to the number of sites in the lattice $M$. So the input layer in  \cref{fig:Algorithm-Predictive-DeepNeuralNetwork} has $M$ number of neurons, and $M$ depends on the lattice size. The number of neurons in the hidden layer is not fixed and depends on the neural network's performance; thus, they are optimized during the neural network optimization using validation matrices. Also, the number of hidden layers is not fixed, and one can use any number of hidden layers, keeping in mind that neural networks should not be under-fitted or over-fitted. The connection between layers is called weights, depicted as lines in \cref{fig:Algorithm-Predictive-DeepNeuralNetwork}. The weights between the input and the first hidden layer are denoted by $W^{(1)}$, the first hidden and the second hidden layer are denoted by $W^{(2)}$ and so on. The weights are the parameters that describe how strongly two neurons are connected. The weights between two layers form a two-dimensional matrix represented as $W^{(r)}_{i,j}$. Each neuron in every layer has its own bias factor $b$ except the input layer, known as ``biases". These weights $W$ and biases $b$ are called trainable parameters of the neural network, and their values are optimized during the training of the neural network. Also, except the input layer, every neuron in each layer is activated by a nonlinear function called ``Activation Function". For our case, $\tanh(x)$ will serve as an activation function. We have not used any activation function in the output layer, as we have already scaled the output with $\ln(x)$. This modification avoids the complex number calculation in neural network architecture.

The value of each neuron in the input layer is assigned as,
\begin{equation}
    \label{eq:Algorithm-Predictive-Architecture-inputLayer}
    Y^{(0)}_i = | \bm{n} \rangle = \begin{bmatrix} n_1 \\ n_2 \\ \vdots \\ n_M \end{bmatrix}
\end{equation}
where, $Y^{(0)}_i$ is the value of each neuron in input layer with $i$ being the index over each neuron in input layer.

The value of each neuron in the first hidden layer is calculated as,
\begin{equation}
    \label{eq:Algorithm-Predictive-Architecture-firstHiddenLayer}
    Y^{(1)}_j = \sum_{i=1}^{M} W_{i,j}^{(1)} Y^{(0)}_i + b_j^{(1)}
\end{equation}
where $Y^{(1)}_j$ is the value of each neuron in the first hidden layer, and $j$ is the index over each neuron in the first hidden layer. $W_{i,j}^{(1)}$ is the weight between the input and the first hidden layer, which is $M \times R_1$ matrix, while $R_1$ is the number of neurons in the first hidden layer. $ b_j^{(1)}$ is the bias factor of the first hidden layer, which is $R_1$-component vector.

In general, the value of each neuron in $n^{th}$ hidden layer, where $1 \leq n \leq L-1$, is give as,

\begin{equation}
    \label{eq:Algorithm-Predictive-Architecture-HiddenLayer}
    Y_j^{(n+1)} = \sum_{i=1}^{R_{(n)}} W_{i,j}^{(n+1)} \mathit{f}(Y^{(n)}_i) + b_j^{(n+1)}
\end{equation}
where, $Y_j^{(n+1)}$ is the value of each neuron in $n^{th}$ hidden layer, $i$ is the index over $(n-1)^{th}$ hidden layer, which given as $i = [1,R_{(n)}]$. $R_{(n)}$ is number of neurons in $(n-1)^{th}$ hidden layer. $b_j^{(n+1)}$ is the bias factor of $n^{th}$ hidden layer and $W_{i,j}^{(n+1)}$ are the weights between $n^{th}$ and $(n-1)^{th}$ hidden layer. $\mathit{f}$ is a nonlinear function that is adopted as,
\begin{equation}
    \label{eq:Algorithm-Predictive-Architecture-activationfun}
    \mathit{f}(x) = \tanh(x)
\end{equation}

Finally, the values of each neuron in the output layer are given as,
\begin{equation}
    \label{eq:Algorithm-Predictive-Architecture-outputLayer}
    Y_k^{(L)} = \sum_{i=1}^{R_{(L-1)}} W_{i,j}^{(L-1)} \mathit{f}(Y^{(L-1)}_i) + b_j^{(L-1)}
\end{equation}
where $Y_k^{(L)}$ value of each neuron in the output layer. The output layer has only two neurons, $k = [1,2]$.

The total number of trainable parameters for this fully connected feed-forward neural network is,
\begin{equation}
    \label{eq:Algorithm-Predictive-Architecture-NoParameters}
    \text{Params}_{\text{ANN}} = \sum_{n=1}^{L-1} R_{(n)}( R_{(n-1)} + 1) + 2R_{(L-1)}
\end{equation}

The complex number $\phi(\bm{n})$ is given as,
\begin{equation}
    \label{eq:Algorithm-Predictive-Architecture-phi}
    \phi(\bm{n}) = \exp{\left[Y_1^{(L)} + i Y_2^{(L)} \right]}
\end{equation}

By calculating $\phi(\bm{n})$ for all possible $| \bm{n} \rangle$, we can construct the quantum state in \cref{eq:Algorithm-Quantum_System-FockState}. With this formalism, the information of the many-body quantum state is stored in the neural network parameters $W$ and $b$. 

To get the information stored correctly, the neural network parameters must be optimized so that the many-body quantum state is close to the ground state. In our case, we will use the concept of reinforcement learning, which is a subset of unsupervised learning, to optimize the neural network for a given cost function $f_c$.

The general formula for cost function $f_c$ using mean squared error is given as,
\begin{equation}
     \label{eq:Algorithm-Predictive-Optimization-costfunctionGen}
     f_c = \frac{1}{n} \sum_{i=1}^n (Y - Y_P)
\end{equation}
where $n$ is the number of data points. Similarly, we can construct cost functions using other matrices. 

Here we are simulating a many-body quantum system. During this simulation, our goal is to optimize the neural network parameters in such a way that the many-body quantum state is close to the ground state, and we are starting from a trial many-body quantum state $|\Psi(\bm{n}) \rangle$ given by \cref{eq:Algorithm-Quantum_System-FockState}. With this setting, we can use a fundamental principle of quantum mechanics to construct the cost function, namely the variational principle. The variational principle states that the expectation value of Hamiltonian for a trail wavefunction will always be higher or equal to the exact ground state energy of the system, which is mathematically represented as
\begin{equation}
\label{eq:Algorithm-Predictive-Optimization-variational}
    E = \langle H \rangle = \frac{\langle \Psi | H | \Psi \rangle}{\langle \Psi | \Psi \rangle} \geq E_0
\end{equation}
where $| \Psi \rangle$ and $E_0$ are the trail wavefunction and the ground state energy, respectively.

Given that starting with a random occupation state $| \bm{n} \rangle$ to construct a trail many-body quantum state $| \Psi(\bm{n}) \rangle$ the expectation value of Hamiltonian given by \cref{eq:Algorithm-Quantum_System-ExpectationValue} will be higher or equal to the expectation value of the ground state wavefunction. With this setting, the expectation value of Hamiltonian given by \cref{eq:Algorithm-Quantum_System-ExpValModified} can be used as a cost function $f_c$ to optimize the neural network. We construct the cost function for the optimization of the deep neural network as,
\begin{equation}
    \label{eq:Algorithm-Predictive-Optimization-costFunction}
    f_c = \langle \Tilde{H} \rangle_M = \left< \sum_{\bm{n'}} \langle \bm{n} | \hat{H} | \bm{n'} \rangle \frac{\phi(\bm{n'})}{\phi(\bm{n})} \right>_{N_T}
\end{equation}
where $N_T$ is the total number of possible states , $\sum_{n'}$ is summation over the sampled states, and $\left< \ldots \right>_{N_T}$ is the average over all possible states.

The optimization of the trainable parameters (weights $W$ and biases $b$) requires the derivative of the cost function against trainable parameters.

The derivative of cost function $f_c$ is given as,
\begin{align}
    \label{eq:Algorithm-Predictive-Optimization-DcostFunction}
    \frac{\partial f_c}{\partial w} &=\frac{\partial \langle \hat{H} \rangle}{\partial w} \nonumber \\
                                    &= \frac{\sum_{\bm{n}, \bm{n'}} \langle \bm{n} | \hat{H} | \bm{n'} \rangle \left[ O_w(\bm{n'}) + O_w^*(\bm{n}) \right] \phi^*(\bm{n}) \phi(\bm{n'})}{\sum_{\bm{n}} |\phi(\bm{n})|^2} \nonumber \\
                                    &- \langle \hat{H} \rangle \frac{\sum_{\bm{n}}  \left[ O_w(\bm{n}) + O_w^*(\bm{n}) |\phi(\bm{n})|^2 \right]}{ \sum_{\bm{n}} |\phi(\bm{n})|^2}  \\
                                    &\simeq 2 {\rm Re} \left( \langle O_w^* \tilde H \rangle_{N_T} - \langle O_w^* \rangle_{N_T} \langle \tilde H \rangle_{N_T} \right) \nonumber
\end{align}
where $O_w(\bm{n'})$ is an operator, given as $O_w(\bm{n'}) = \frac{1}{\phi(\bm{n'})} \frac{\partial \phi(\bm{n'})}{\partial w}$.

In the derivative of the cost function, we have introduced an operator $O_w^*(\bm{n})$, which is given as $O_w^*(\bm{n}) = \frac{1}{\phi^*(\bm{n})} \frac{\partial \phi^*(\bm{n})}{\partial w}$. We also need to compute this operator as it is given as derivative of complex number $\phi^*(\bm{n})$. From \cref{eq:Algorithm-Predictive-Architecture-phi}, we will use the value of $\phi^*(\bm{n})$ to compute the operator. The calculation of this operator is given as,
\begin{align}
    \label{eq:Algorithm-Predictive-Optimization-operator}
    O_w^*(\bm{n}) &= \frac{1}{\phi^*(\bm{n})} \frac{\partial \phi^*(\bm{n})}{\partial w} \nonumber \\
                  &= \frac{1}{\phi^*(\bm{n})} \frac{\partial}{\partial w} \left( \exp{\left[Y_1^{ (L)} - i Y_2^{(L)} \right]} \right) \nonumber \\
                  &= \frac{1}{\phi^*(\bm{n})} \exp{\left[Y_1^{ (L)} - i Y_2^{(L)} \right]} \frac{\partial}{\partial w} \left[ Y_1^{(L)} - i Y_2^{(L)} \right] \\
                  &= \frac{\partial Y_1^{ (L)}}{\partial w} - i \frac{\partial Y_2^{ (L)}}{\partial w} \nonumber
\end{align}
Here, $\frac{\partial Y_1^{ (L)}}{\partial w}$ and $\frac{\partial Y_2^{ (L)}}{\partial w}$ can be calculated using chain rule with respect to each weight $W$ and bias $b$ matrix. We will compute these values using this algorithm for a many-body quantum system simulation in \cref{sec:method}.

Using the derivative of the cost function, three optimizers algorithms, namely, the steepest descent optimizer, the adaptive gradient (AdaGrad) optimizer, and the adaptive momentum (Adam) optimizer can be used to optimize the trainable parameter.

The steepest descent optimizer is a widely used optimization technique. A gradient-based optimization algorithm aims to minimize the objective function by taking steps toward the steepest descent. The steepest descent is the direction of the negative gradient of the objective function. The Steepest Descent algorithm works by computing the gradient of the objective function with respect to the model's parameters. It then takes a step in the direction of the negative gradient. The step size is determined by the learning rate, a hyperparameter that controls the update size at each iteration. The advantage of the steepest descent algorithm is its simplicity. It is easy to implement and computationally efficient. However, it suffers from some limitations. The steepest descent algorithm may converge slowly, particularly when the objective function is ill-conditioned. Additionally, the learning rate can be challenging to tune, and if it is set too high, the algorithm may fail to converge.

The neural network parameters using the steepest descent optimizer are updated as follows,
\begin{equation}
    \label{eq:Algorithm-Predictive-Optimization-steep}
    w \rightarrow w - \gamma\frac{\partial \langle H \rangle_{N_T}}{\partial w}
\end{equation}
where $w$ is neural network parameters, the weights $W$ and biases $b$, and $\gamma$ is learning rate while $\gamma < 1$. $\frac{\partial \langle  H \rangle_{N_T}}{\partial w}$ is given by \cref{eq:Algorithm-Predictive-Optimization-DcostFunction}.

For the derivative of the cost function in \cref{eq:Algorithm-Predictive-Optimization-DcostFunction}, we can modify the \cref{eq:Algorithm-Predictive-Optimization-steep} as follows,
\begin{equation}
    \label{eq:Algorithm-Predictive-Optimization-steepMod}
    w \rightarrow w - \gamma 2 {\rm Re} \left( \langle O_w^* H \rangle_{N_T} - \langle O_w^* \rangle_{N_T} \langle \tilde H \rangle_{N_T} \right)
\end{equation}

The adaptive gradient (AdaGrad) optimizer is an optimization technique that addresses some of the limitations of the steepest descent algorithm. The AdaGrad algorithm adapts the learning rate of each parameter based on the historical gradients. The idea behind AdaGrad is to reduce the learning rate for parameters with a large gradient magnitude and to increase it for those with a small gradient magnitude. The AdaGrad algorithm works by accumulating the square of the gradients for each parameter over time. The accumulated gradients are then used to scale the learning rate for each parameter. This means that parameters with large gradients will have a smaller learning rate, while parameters with small gradients will have a larger learning rate. The advantage of the AdaGrad algorithm is that it automatically adapts the learning rate for each parameter. This can lead to faster convergence and better performance, particularly when the objective function is ill-conditioned. However, the AdaGrad algorithm suffers from some limitations. The accumulated gradients can become very large over time, leading to very small learning rates. This can cause the algorithm to converge too slowly or even fail to converge.

The neural network parameters using the adaptive gradient optimizer are updated as follows,
\begin{align}
    \label{eq:Algorithm-Predictive-Optimization-AdaGrad}
    w_i & \rightarrow w_i - \frac{\gamma}{\sqrt{v_i}+\epsilon}\frac{\partial \langle H \rangle}{\partial w_i} \\
    v_i & \rightarrow v_i + \left(\frac{\partial \langle H \rangle}{\partial w_i}\right)^2 \nonumber
\end{align}
where $\gamma$ and $\epsilon$ are the learning rates where $\gamma < 1$ and $\epsilon \ll 1$.  $\frac{\partial \langle \tilde H \rangle_{N_T}}{\partial w}$ is given by \cref{eq:Algorithm-Predictive-Optimization-DcostFunction}

For the derivative of the cost function in \cref{eq:Algorithm-Predictive-Optimization-DcostFunction}, we can modify the \cref{eq:Algorithm-Predictive-Optimization-AdaGrad} as follows,
\begin{align}
    \label{eq:Algorithm-Predictive-Optimization-AdaGradMod}
    w_i & \rightarrow w_i - \frac{\gamma}{\sqrt{v_i}+\epsilon} 2{\rm Re} \left( \langle O_w^* H \rangle_{N_T} - \langle O_w^* \rangle_{N_T} \langle \tilde H \rangle_{N_T} \right) \\
    v_i & \rightarrow v_i + 4\left({\rm Re} \left( \langle O_w^* H \rangle_{N_T} - \langle O_w^* \rangle_{N_T} \langle \tilde H \rangle_{N_T} \right)\right)^2 \nonumber
\end{align}

The adaptive momentum (Adam) optimizer is a modification of the AdaGrad algorithm. The Adam algorithm incorporates momentum into the optimization process by keeping track of the gradients' first and second moments. The first moment is the mean of the gradients, while the second is the gradients' variance. The Adam algorithm uses these moments to compute an adaptive learning rate for each parameter. The Adam algorithm computes each parameter's first and second moments of the gradients. It then uses these moments to compute an adaptive learning rate for each parameter. The algorithm also uses a momentum term, which helps to smooth out the updates and prevent oscillations. The advantage of the Adam algorithm is its ability to adapt the learning rate for each parameter while incorporating momentum. This can lead to faster convergence and better performance, particularly when the objective function is non-convex. The Adam algorithm is also robust to the choice of hyperparameters and has been shown to work well in a wide range of settings. However, the Adam algorithm can suffer from overfitting, particularly when the training data or the model is too noisy.

The neural network parameters using the adaptive momentum optimizer are updated as follows,
\begin{align}
    \label{eq:Algorithm-Predictive-Optimization-Adm}
    w_i & \rightarrow w_i - \gamma \frac{m_i}{1-\beta^l_1}\frac{1}{\sqrt{\frac{v_i}{i-\beta^l_2}}+\epsilon} \nonumber \\
    m_i & \rightarrow \beta_1 m_i + (1-\beta_1)\frac{\partial \langle H \rangle}{\partial w_i} \\
    v_i & \rightarrow \beta_2 v_i + (1-\beta_2) \left(\frac{\partial \langle H \rangle}{\partial w_i}\right)^2 \nonumber
\end{align}
where, $\gamma < 1$ , $\beta_1 = 0.9$ and $\beta_2 = 0.999$ for $l^{th}$ update. The initial values of $v_i$ \& $m_i$ are zero.

For the derivative of the cost function in \cref{eq:Algorithm-Predictive-Optimization-DcostFunction}, we can modify the \cref{eq:Algorithm-Predictive-Optimization-Adm} as follows,
\begin{align}
    \label{eq:Algorithm-Predictive-Optimization-AdmMod}
    w_i & \rightarrow w_i - \gamma \frac{m_i}{1-\beta^l_1}\frac{1}{\sqrt{\frac{v_i}{i-\beta^l_2}}+\epsilon} \nonumber \\
    m_i & \rightarrow \beta_1 m_i + 2(1-\beta_1){\rm Re} \left( \langle O_w^* H \rangle_{N_T} - \langle O_w^* \rangle_{N_T} \langle \tilde H \rangle_{N_T} \right) \\
    v_i & \rightarrow \beta_2 v_i + 4(1-\beta_2) \left({\rm Re} \left( \langle O_w^* H \rangle_{N_T} - \langle O_w^* \rangle_{N_T} \langle \tilde H \rangle_{N_T} \right)\right)^2 \nonumber
\end{align}
%---------------------------------------------------------------------------------
%============================

% ========== METHOD ==========
\section{\label{sec:method}Method of calculations}
\begin{figure}
    \centering
    \includegraphics[scale = 0.35]{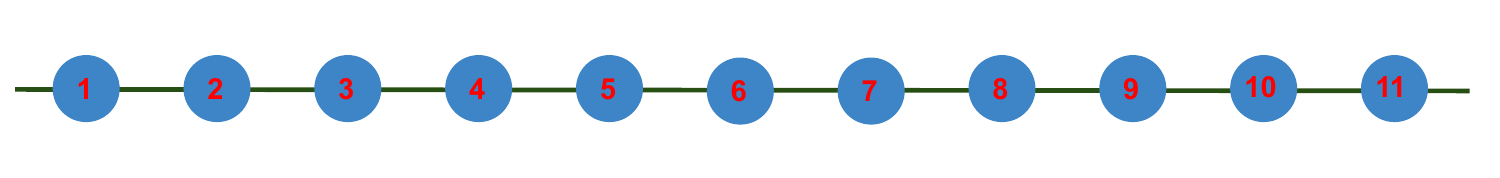}
    \caption{One dimensional lattice chain with number of sites $M = 11$.}
    \label{fig:Method-Lattice}
\end{figure}

Let us consider a one-dimensional chain of the lattice with the number of sites $M = 11$, depicted in \cref{fig:Method-Lattice} with the total number of bosons $N = 9$. 

%----- The Neural network --------------------------------------------------------
\subsection{\label{subsec:Method-Architecture} The Deep Neural Network Architecture}
\begin{figure}
    \centering
    \includegraphics[scale = 0.4]{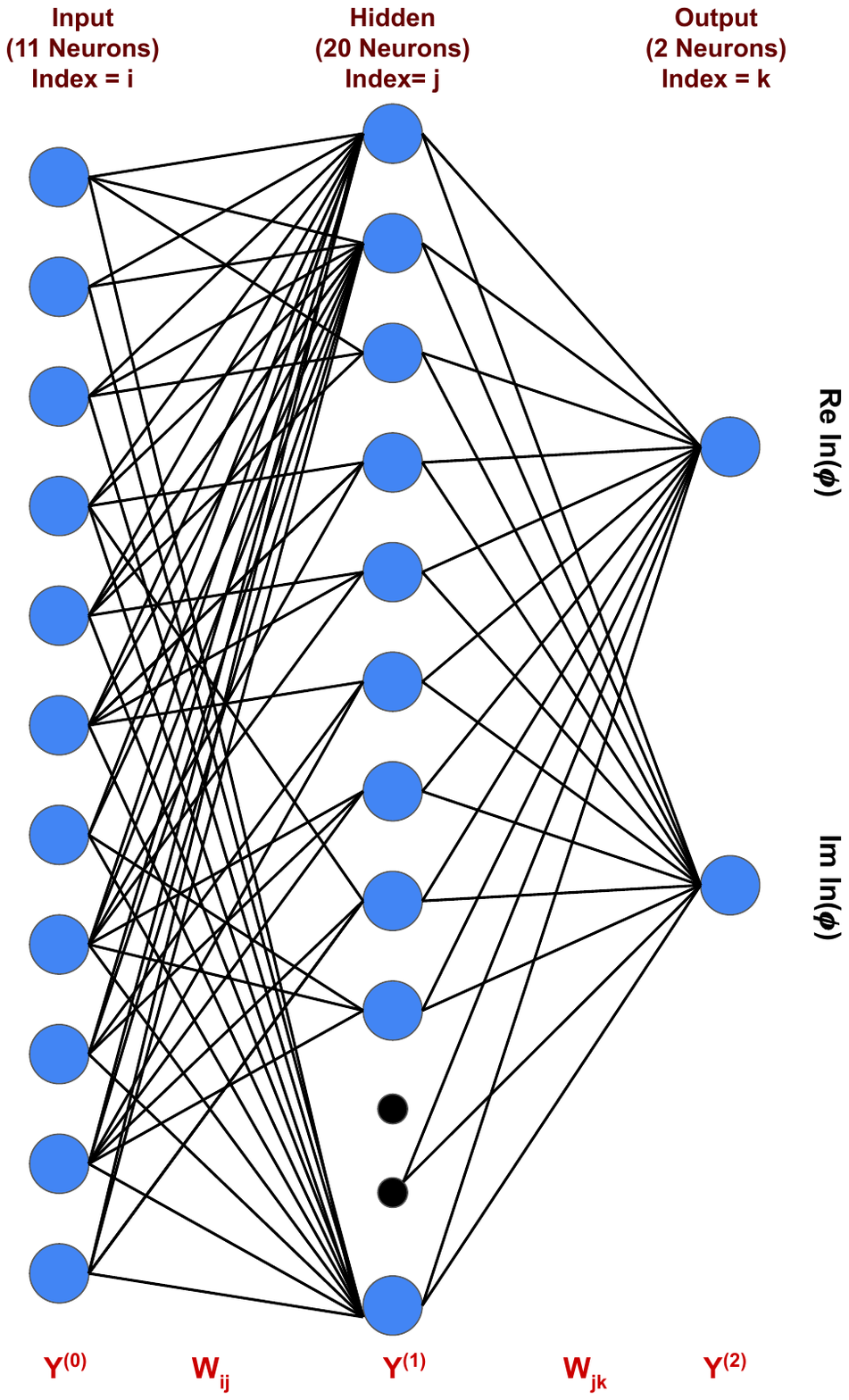}
    \caption{The deep neural network architecture to simulate the bosonic quantum system.}
    \label{fig:Method-Architecture-NeuralNetwork}
\end{figure}

\cref{fig:Method-Architecture-NeuralNetwork} depicts the deep neural network architecture used to simulate the bosonic system in a one-dimensional lattice chain with one input, one hidden, and one output layer. The input layer has $11$ neurons as we have $11$ sites in the lattice, the hidden layer has $20$ neurons, and the output layer has $2$ neurons. 

The value of each neuron in the input layer is given as,
\begin{equation}
    \label{eq:Method-Architecture-Input}
    Y_i^{(0)} = | n_i \rangle
\end{equation}

The value of each neuron in the hidden layer is calculated as,
\begin{equation}
    \label{eq:Method-Architecture-Hidden}
    Y_j^{(1)} = \sum_{i=1}^M W_{ij} Y_i^{(0)} + b_j
\end{equation}
where $i$, $j$, $W_{ij}$, and $b_j$ are the indices over the input and the hidden layers, weights between the input and the hidden layers, and the bias for the hidden layer, respectively.

The value of each neuron in the output layer is calculated as,
\begin{equation}
    \label{eq:Method-Architecture-Optput}
    Y_k^{(2)} = \sum_{j=1}^{20} W_{jk} f(Y_j^{(1)})
\end{equation}
where $k$ is the index over the output layer, $W_{jk}$ are the weights between the output and hidden layer, and $f(x)$ is the nonlinear function of the hidden layer, which is given as $f(x) = \tanh(x)$.

The \cref{eq:Method-Architecture-Input},  \cref{eq:Method-Architecture-Hidden}, and  \cref{eq:Method-Architecture-Optput} defines the architecture of the neural network, which are formulated from \cref{eq:Algorithm-Predictive-Architecture-HiddenLayer} for the neural network depicted in \cref{fig:Method-Architecture-NeuralNetwork}.
%---------------------------------------------------------------------------------

%----- The quantum Many-Body States ----------------------------------------------
\subsection{\label{subsec:Method-Quantum} The quantum Many-Body States}
The many-body quantum state for the bosons in lattice with $M = 11$ sites, using \cref{eq:Algorithm-Quantum_System-FockState}, is given as,
\begin{equation}
    \label{eq:Method-Quantum-State}
    | \Psi \rangle = \sum_{i = 1}^M \phi(n_1, n_2, \ldots, n_{11}) | n_1, n_2, \ldots, n_{11} \rangle
\end{equation}

For $9$ bosons on $11$ lattice sites, the total number of possible states can be calculated using \cref{eq:Algorithm-Quantum_System-Total_Fock_States} as
\begin{align}
    \label{eq:Method-Quantum-TotalStates}
    N_\text{tot} &= \frac{(N + M -1)!}{N! (M-1)!} \nonumber \\
                 &= \frac{(9 + 11 -1)!}{9! (11-1)!} \\
                 &\simeq 10^6. \nonumber
\end{align}

The complex coefficient $\phi(n_1, n_2, ..., n_{11})$ in \cref{eq:Method-Quantum-State} is calculated using the values of the output layer of the neural network given in \cref{eq:Method-Architecture-Optput} as
\begin{equation}
    \label{eq:Method-Quantum-Phi}
    \phi(n_1, n_2, ..., n_{11}) = \exp{Y_1^{(2)} + i Y_2^{(2)}}
\end{equation}

To simulate the bosonic system, we have considered the Bose-Hubbard model, as described by the Hamiltonian
\begin{equation}
    \label{eq:Method-Quantum-Hamiltonian}
    H = -J \sum_{\langle i j \rangle} \hat a_i^\dagger \hat a_j+ \sum_i \frac{U}{2} \hat n_i (\hat n_i - 1),
\end{equation}
where $J > 0$ is hopping coefficient and $U$ is the on-site interaction energy. $\hat a_i^\dagger$ is a creation operator which creates a boson on $i^{th}$ site and $\hat a_j$ is annihilation operator which annihilates a boson at $j^{th}$ site. $\hat n_i$ is the number operator, where the commutation relation $[\hat a_i^\dagger, \hat a_j] = \delta_{i,j}$ is satisfied. The expectation value of this Hamiltonian can be calculated using \cref{eq:Algorithm-Quantum_System-ExpValModified}.

The expectation value of Hamiltonian, given in \cref{eq:Method-Quantum-Hamiltonian}, is calculated for a stochastic process as
\begin{align}
    \label{eq:Method-Quantum-EpxectModified}
    \langle \Tilde{H} \rangle =  \left< \sum_{\bm{n'}} \langle \bm{n} | \hat{H} | \bm{n'} \rangle \frac{\phi(\bm{n'})}{\phi(\bm{n})} \right>_{N_T},
\end{align}
where $\phi(\bm{n'})$ and $\phi(\bm{n})$ are calculated using \cref{eq:Method-Quantum-Phi}. $N_T$ is the total number of possible states. For Bose-Hubbard Hamiltonian give by \cref{eq:Method-Quantum-Hamiltonian}, we can compute the value of $\langle \bm{n} | \hat{H} | \bm{n'} \rangle$.

The value of $\langle \bm{n} | \hat{H} | \bm{n'} \rangle$ for Bose-Hubbard Hamiltonian is calculated as,
\begin{align}
    \label{eq:Method-Quantum-<n|H|n'>}
    &\langle \bm{n} | \hat{H} | \bm{n'} \rangle = \langle \bm{n} | -J \sum_{\langle i j \rangle} \hat a_i^\dagger \hat a_j+ \sum_i \frac{U}{2} \hat n_i (\hat n_i - 1) | \bm{n'} \rangle \nonumber \\
                                       &= -J\langle \bm{n} | \sum_{\langle i j \rangle} \hat a_i^\dagger \hat a_j| \bm{n'} \rangle + \frac{U}{2} \langle \bm{n} | \sum_i \hat n_i (\hat n_i - 1) | \bm{n'} \rangle  \\
                                       &= -J \left[ \sum_{i=1}^{n_i} \sqrt{n_i + 1} \left( \sqrt{n_{i-1}} \delta_{\bm{n_i}, \bm{n_{i-1}}} + \sqrt{n_{i+1}} \delta_{\bm{n_i}, \bm{n_{i+1}}} \right) \right] \nonumber \\
                                       & + \frac{U}{2} \left[ \sum_i n_i(n_i-1) \delta_{n,n'} \right] \nonumber
\end{align}
where $n_i$ is the number of bosons at $i^{th}$ site of the lattice. $\delta_{\bm{n_i}, \bm{n_{i-1}}}$ and $\delta_{\bm{n_i}, \bm{n_{i+1}}}$ are give as follows,
\begin{align}
    \label{eq:Method-Quantum-delta}
    \delta_{\bm{n_i}, \bm{n_{i-1}}} &= \langle \bm{n} | n_1, n_2, ..., n_{i-1}-1, n_i+1, ..., n_{11} \rangle \\
    \delta_{\bm{n_i}, \bm{n_{i+1}}} &= \langle \bm{n} | n_1, n_2, ..., n_{i}+1, n_{i+1}-1, ..., n_{11} \rangle \nonumber
\end{align}
%---------------------------------------------------------------------------------

%---- The optimization of the neural network -------------------------------------
\subsection{\label{subsec:Method-Optimization} The Optimization Of The Neural Network}
The derivative of the cost function, given by \cref{eq:Method-Quantum-EpxectModified}, is computed using \cref{eq:Algorithm-Predictive-Optimization-DcostFunction}, and given as,
\begin{equation}
    \label{eq:Method-Optimization-Dcost}
    \frac{\partial \langle H \rangle}{\partial w} =  2 {\rm Re} \left( \langle O_w^* \tilde H \rangle_{N_T} - \langle O_w^* \rangle_{N_T} \langle \tilde H \rangle_{N_T} \right)
\end{equation}
where $O_w^*$ is an operator given as,
\begin{align}
    \label{eq:Method-Quantum-operator}
    O_w^* = \frac{\partial Y_1^{(2)}}{\partial w} - i \frac{\partial Y_2^{(2)}}{\partial w}
\end{align}
where $Y_1^{(2)}$ and $Y_2^{(2)}$ is output of the neurons of output layer given by \cref{eq:Method-Architecture-Optput}.

Given that we have only three layers in the neural network, we can compute the derivatives of $Y_1^{(2)}$ and $Y_2^{(2)}$ with respect to network parameters. The derivative of $Y_1^{(2)}$ and $Y_2^{(2)}$ with respect to $w = W_{jk}$ is,
\begin{align}
    \label{eq:Method-Optimization-dY1Y2Wjk}
    \frac{\partial Y_1^{(2)}}{\partial W_{jk}} &= \frac{\partial}{\partial W_{jk}} \sum_{j=1}^{20} W_{jk} \tanh(Y_j^{(1)}) \nonumber \\ 
    & = \sum_{j=1}^{20} \tanh(Y_j^{(1)}) \\
    \frac{\partial Y_2^{(2)}}{\partial W_{jk}} &= \frac{\partial}{\partial W_{jk}} \sum_{j=1}^{20} W_{jk} \tanh(Y_j^{(1)}) \nonumber \\
    &= \sum_{j=1}^{20} \tanh(Y_j^{(1)})
\end{align}

The derivative of $Y_1^{(2)}$ and $Y_2^{(2)}$ with respect to $w = W_{ij}$ is,
\begin{align}
    \label{eq:Optimization-Quantum-dY1Y2Wij}
    \frac{\partial Y_1^{(2)}}{\partial W_{ij}} &= \frac{\partial}{\partial W_{ij}} \sum_{j=1}^{20} W_{jk} \tanh(Y_j^{(1)}) \nonumber \\
    &= N  \sum_{j=1}^{20} W_{jk} sech^2(Y_j^{(1)}) \\ 
    \frac{\partial Y_2^{(2)}}{\partial W_{ij}} &= \frac{\partial}{\partial W_{ij}} \sum_{j=1}^{20} W_{jk} \tanh(Y_j^{(1)}) \nonumber \\
    &= N  \sum_{j=1}^{20} W_{jk} sech^2(Y_j^{(1)})
\end{align}
Where $N$ is the total number of bosons in the system. Similarly, we can compute the derivative of $Y_1^{(2)}$ and $Y_2^{(2)}$ with respect to $w = b_j$.

By putting these derivatives in \cref{eq:Method-Quantum-operator}, we can compute the final value of $O_w^*$, which is as follows,
\begin{equation}
    \label{eq:Method-Optimization-operatorFinal}
    O_w^* = 
            \begin{cases}
                N \sum_{j=1}^{20} sech^2(Y_j^{(1)}) (1-i) \ \text{if } w = W_{ij} \\
                \sum_{j=1}^{20} \tanh(Y_j^{(1)}) (1-i) \ \ \ \ \ \text{if } w = W_{jk} \\
                \sum_{j=1}^{20} sech^2(Y_j^{(1)}) (1-i) \ \ \ \ \text{if } w = b_{j}
            \end{cases}
\end{equation}

With this, we have constructed the derivative of the cost function.

The steepest descent optimizer, given in \cref{eq:Algorithm-Predictive-Optimization-steep}, can be expanded for each weight and bias matrix as follows,
\begin{align}
    \label{eq:Method-Optimization-steep}
    W_{jk} &\rightarrow W_{jk} - \gamma 2 {\rm Re}\left( \langle O^*_{W_{jk}} \tilde H \rangle_{N_T} - \langle O^*_{W_{jk}} \rangle_{N_T} \langle \tilde H \rangle_{N_T} \right) \nonumber \\
    W_{ij} &\rightarrow W_{ij} - \gamma 2 {\rm Re}\left( \langle O^*_{W_{ij}} \tilde H \rangle_{N_T} - \langle O^*_{W_{ij}} \rangle_{N_T} \langle \tilde H \rangle_{N_T} \right) \nonumber \\
    b_{j} &\rightarrow b_{j} - \gamma 2 {\rm Re}\left( \langle O^*_{b_{j}} \tilde H \rangle_{N_T} - \langle O^*_{b_{j}} \rangle_{N_T} \langle \tilde H \rangle_{N_T} \right) 
\end{align}
where, $O^*_{W_{jk}}$, $O^*_{W_{ij}}$, and $O^*_{b_{j}}$ are given by \cref{eq:Method-Optimization-operatorFinal}. Similarly, we can expand the \cref{eq:Algorithm-Predictive-Optimization-AdmMod}, and \cref{eq:Algorithm-Predictive-Optimization-AdaGradMod} for adaptive momentum and adaptive gradient optimizers, respectively.

% =============================================
% ========== RESULTS AND DISCUSSIONS ==========
\section{\label{sec:result}Results and Discussions}
After developing the machine learning algorithm to solve for a bosonic many-body quantum system, we test the algorithm for its usefulness. To test the artificial intelligence-driven quantum many-body algorithm, we have solved the Bose-Hubbard Hamiltonian using the developed algorithm through the method presented in \cref{sec:method}. This section will discuss the results obtained for the Bose-Hubbard Hamiltonian on a one-dimensional lattice chain. The main motive of the results is to validate the intelligence-driven algorithm for the bosonic system, so we will only compute the ground state energy of the quantum system. We will use the cost function optimization graph to explain the algorithm's accuracy. Also, we will validate the optimizers to find the best working optimizer by presenting the reasons behind them.

\begin{figure}
    \centering
    \includegraphics[scale = 0.7]{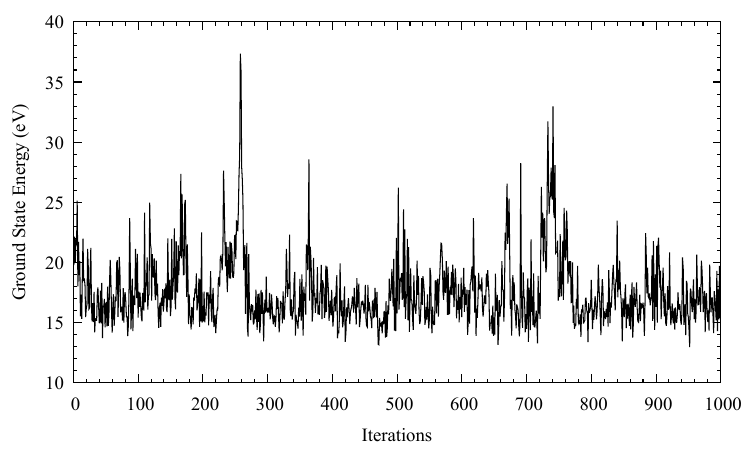}
    \caption{The graph between ground state energy (in eV) of the bosonic many-body quantum system with respect to the iterations for the steepest descent optimizer using the learning rate $\gamma = 0.01$.}
    \label{fig:Results-Steep}
\end{figure}

\begin{figure}
    \centering
    \includegraphics[scale = 0.7]{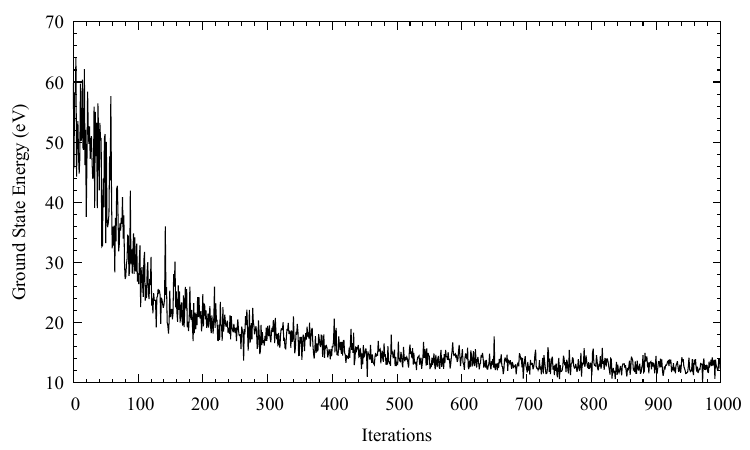}
    \caption{The graph between ground state energy (in eV) of the bosonic many-body quantum system with respect to the iterations for the adaptive gradient optimizer using the learning rate $\gamma = 0.01$ and $\epsilon = 0.99$.}
    \label{fig:Results-AdaGrad}
\end{figure}

\begin{figure}
    \centering
    \includegraphics[scale = 0.7]{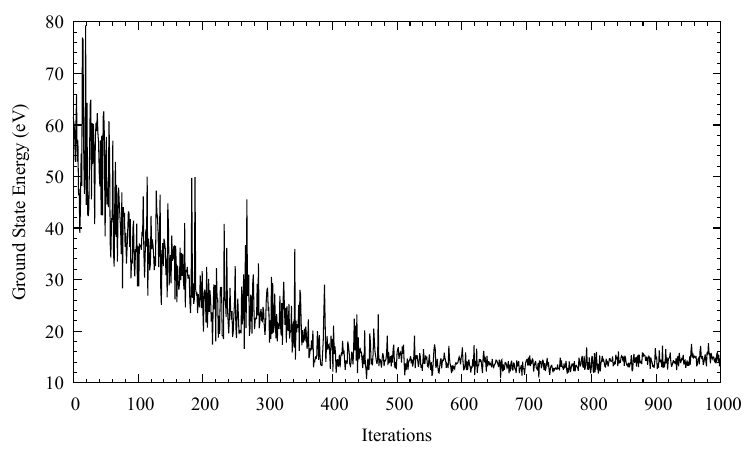}
    \caption{The graph between ground state energy (in eV) of the bosonic many-body quantum system with respect to the iterations for the adaptive momentum optimizer using the learning rate $\gamma = 0.01$, $\beta_1 = 0.9$, and $\beta_2 = 0.999$.}
    \label{fig:Results-Adm}
\end{figure}

To simulate the Bose-Hubbard model in a one-dimensional lattice chain, we have taken a 1D lattice with $M = 11$ sites occupied by $N = 9$ bosons. Also, the simulation is done taking the values of hopping term $J = 1$~eV and onsite interaction term $U = 3$~eV for the Hamiltonian given in \cref{eq:Method-Quantum-Hamiltonian}. We have performed the simulation for $10^3$ iterations to achieve convergence. However, our results show that $10^3$ iterations are not required to achieve the convergence; the neural network converges with fully optimized neural network parameters way before that. To compute the expectation value of Hamiltonian (or cost function) and derivative of the cost function in each interaction, we have sampled $10^3$ quantum states using the method discussed in \cref{subsec:Method-Quantum}, and \cref{subsec:Method-Optimization}. Then the expectation value of the Hamiltonian (or the cost function) with respect to each iteration is plotted in \cref{fig:Results-Steep}, \cref{fig:Results-AdaGrad}, and \cref{fig:Results-Adm} for steepest descent, adaptive gradient, and adaptive momentum optimizers, respectively.

\cref{fig:Results-Steep} depicts the graph between ground state energy (in eV) of the bosonic many-body quantum system with respect to the iterations for the steepest descent optimizer using the learning rate $\gamma = 0.01$, which shows that the neural network had not appropriately converged. As previously discussed, the steepest descent optimizer is the simplest optimization algorithm; thus, it suffers from some limitations. This algorithm uses a fixed learning rate, which makes the optimization of the neural network difficult, especially if the cost function is complicated. Here, the cost function is given by the expectation value of the Hamiltonian; thus, the optimization algorithm is facing a problem in optimizing the neural network parameters. The choppiness in the plot shows that the optimization algorithm has stuck into the local minima of the cost function and cannot overcome it. One way to solve this problem is to increase the learning rate, so we have tried the simulation with a higher learning rate $\gamma = 0.1$. The higher learning rate helped the optimization algorithm overcome the local minima. However, then due to the high learning rate, the optimization algorithm missed the global minima and overshot it. With that, we again got the same pattern in the graph. So, with these results for the steepest descent optimizer, we can gather that it is not a good algorithm to optimize the neural network for a bosonic many-body quantum system.

After using the steepest descent optimizer, we used the adaptive gradient optimizer to train the neural network to overcome the challenges faced using the previous optimizer. The \cref{fig:Results-AdaGrad} depicts the graph between ground state energy (in eV) of the bosonic many-body quantum system with respect to the iterations for the adaptive gradient optimizer using the learning rate $\gamma = 0.01$ and $\epsilon = 0.99$. The graph shows that the neural network is adequately trained, and the neural network parameters are efficiently adjusted. This was achieved as the learning rate during optimization changes as we are close to the global minima; this helps overcome the local minima problem. However, when we are close to the global minima, the learning rate is small enough not to overshoot it. The cost function value starts from $\sim 60$~eV and slowly converges to $\sim 12$~eV. We ran this simulation for $10^3$ iterations, but the neural network almost converged by $500$. However, the adaptive gradient optimizer was able to optimize the neural network, but we have previously discussed this optimizer has certain limitations. If accumulated gradients become very large, the learning rate becomes very slow, and the convergence suffers. We have not faced this problem for our case as we have used $J = 1$~eV and $U = 3$~eV. Nevertheless, accumulated gradients can become very large if the values of $J$ and $U$ are high because the system's energy becomes large, which is used as a cost function, making gradients very large.

\begin{table}
    \centering
    \begin{tabular}{c|c}
    \hline
    \hline
        \textbf{Optimizer} & \textbf{Ground State Energy ($\varepsilon$~eV)} \\
    \hline
        Steepest Descent & 17.8 \\
        Adaptive Gradient & 10.6 \\
        Adaptive Momentum & 12.5 \\
    \hline
    \end{tabular}
    \caption{The table for the ground state energies obtained using the optimizer algorithms.}
    \label{tab:Results-GroundEnergy}
\end{table}

In the case of large accumulated gradients, the adaptive gradient is no longer an excellent algorithm to optimize the neural network. To avoid this problem, we have developed another optimizer called ``Adaptive Momentum Optimizer". The \cref{fig:Results-Adm} depicts the graph between ground state energy (in eV) of the bosonic many-body quantum system with respect to the iterations for the adaptive momentum optimizer using the learning rate $\gamma = 0.01$, $\beta_1 = 0.9$, and $\beta_2 = 0.999$. The graph shows that the neural network is adequately trained, and the neural network parameters are efficiently adjusted. As we discussed, the adaptive momentum optimizer optimizes the neural network parameters by computing each parameter's first and second moments of the gradient. This allows the algorithm to train the neural network efficiently even if the gradients become large. Also, the gradient moments help achieve faster neural network convergence. Also, the algorithm can optimize the neural network parameters if the cost function is non-convex. However, in our case, the cost function will never become non-convex. The \cref{tab:Results-GroundEnergy} shows the energies obtained using the optimizers used in this algorithm.
% ====================
% ========== CONCLUSION ==========
\section{\label{sec:conc}Conclusion}
The primary objective of this study was to devise an algorithm employing cutting-edge deep neural networks to address many-body quantum systems. This work delineates the algorithm's structure, divided into three components: quantum system, sampling algorithm, and predictive algorithm. Before delving into mathematical formalism, we elucidated each part's utility, considering lattice shape and size adaptability. Notably, our algorithm surpasses traditional methods by accommodating frustrated lattices and simulating large systems, which are typically challenging for methods like quantum Monte Carlo or exact diagonalization.

Using combination theory, we defined the quantum state using many-body formalism and computed the total number of possible quantum states. Further, we detailed the construction of the entire quantum state and calculated the Hamiltonian's expectation value. Subsequently, we developed sampling laws, including local sampler and exchange sampler, and associated probability calculations for accepting or rejecting proposed quantum states. Due to the stochastic nature of computations, we modified the expectation value of the Hamiltonian accordingly.

The pivotal aspect, the predictive algorithm, was elaborated here. We devised a neural network architecture and outlined its operation in predicting quantum states. We adopted a reinforcement learning approach for optimization, formulating the cost function based on the variational principle. Three optimizers were proposed: steepest descent, adaptive gradient (AdaGrad), and adaptive momentum (Adam), each tailored to mitigate specific limitations.

Next, we demonstrated our method's efficacy by solving the Bose-Hubbard model on a one-dimensional lattice chain. The neural network's architecture, expectation value computation, and optimization procedures were detailed. Despite initial challenges with the steepest descent optimizer, adaptive gradient optimization successfully achieved the many-body ground state with an energy of approximately $10.6$ eV. We highlighted adaptive momentum optimization's utility in handling large gradient accumulations.

Overall, this algorithm represents a significant advancement in simulating quantum systems, offering capabilities beyond traditional methods.
% ====================
% ========== ACKNOWLEDGMENT ==========
\begin{acknowledgments}
Financial support from SERB India through grant numbers ECR/2016/001004, CRG/2021/005320, DST India INSPIRE fellowship through grant number IF171000, and the use of the high-performance computing facility at IISER Bhopal are gratefully acknowledged.
\end{acknowledgments}
% ====================
% ========== BIBLIOGRAPHY ==========
%\bibliography{library}

\begin{thebibliography}{17}%
\makeatletter
\providecommand \@ifxundefined [1]{%
 \@ifx{#1\undefined}
}%
\providecommand \@ifnum [1]{%
 \ifnum #1\expandafter \@firstoftwo
 \else \expandafter \@secondoftwo
 \fi
}%
\providecommand \@ifx [1]{%
 \ifx #1\expandafter \@firstoftwo
 \else \expandafter \@secondoftwo
 \fi
}%
\providecommand \natexlab [1]{#1}%
\providecommand \enquote  [1]{``#1''}%
\providecommand \bibnamefont  [1]{#1}%
\providecommand \bibfnamefont [1]{#1}%
\providecommand \citenamefont [1]{#1}%
\providecommand \href@noop [0]{\@secondoftwo}%
\providecommand \href [0]{\begingroup \@sanitize@url \@href}%
\providecommand \@href[1]{\@@startlink{#1}\@@href}%
\providecommand \@@href[1]{\endgroup#1\@@endlink}%
\providecommand \@sanitize@url [0]{\catcode `\\12\catcode `\$12\catcode
  `\&12\catcode `\#12\catcode `\^12\catcode `\_12\catcode `\%12\relax}%
\providecommand \@@startlink[1]{}%
\providecommand \@@endlink[0]{}%
\providecommand \url  [0]{\begingroup\@sanitize@url \@url }%
\providecommand \@url [1]{\endgroup\@href {#1}{\urlprefix }}%
\providecommand \urlprefix  [0]{URL }%
\providecommand \Eprint [0]{\href }%
\providecommand \doibase [0]{https://doi.org/}%
\providecommand \selectlanguage [0]{\@gobble}%
\providecommand \bibinfo  [0]{\@secondoftwo}%
\providecommand \bibfield  [0]{\@secondoftwo}%
\providecommand \translation [1]{[#1]}%
\providecommand \BibitemOpen [0]{}%
\providecommand \bibitemStop [0]{}%
\providecommand \bibitemNoStop [0]{.\EOS\space}%
\providecommand \EOS [0]{\spacefactor3000\relax}%
\providecommand \BibitemShut  [1]{\csname bibitem#1\endcsname}%
\let\auto@bib@innerbib\@empty
%</preamble>
\bibitem [{\citenamefont {Georges}\ \emph {et~al.}(1996)\citenamefont
  {Georges}, \citenamefont {Kotliar}, \citenamefont {Krauth},\ and\
  \citenamefont {Rozenberg}}]{antoine1996}%
  \BibitemOpen
  \bibfield  {author} {\bibinfo {author} {\bibfnamefont {A.}~\bibnamefont
  {Georges}}, \bibinfo {author} {\bibfnamefont {G.}~\bibnamefont {Kotliar}},
  \bibinfo {author} {\bibfnamefont {W.}~\bibnamefont {Krauth}},\ and\ \bibinfo
  {author} {\bibfnamefont {M.~J.}\ \bibnamefont {Rozenberg}},\ }\bibfield
  {title} {\bibinfo {title} {Dynamical mean-field theory of strongly correlated
  fermion systems and the limit of infinite dimensions},\ }\href
  {https://10.1103/RevModPhys.68.13} {\bibfield  {journal} {\bibinfo  {journal}
  {Rev. Mod. Phys.}\ }\textbf {\bibinfo {volume} {68}},\ \bibinfo {pages} {13}
  (\bibinfo {year} {1996})}\BibitemShut {NoStop}%
\bibitem [{\citenamefont {Schollw\"ock}(2005)}]{Schollwock2005}%
  \BibitemOpen
  \bibfield  {author} {\bibinfo {author} {\bibfnamefont {U.}~\bibnamefont
  {Schollw\"ock}},\ }\bibfield  {title} {\bibinfo {title} {The density-matrix
  renormalization group},\ }\href {https://10.1103/RevModPhys.77.259}
  {\bibfield  {journal} {\bibinfo  {journal} {Rev. Mod. Phys.}\ }\textbf
  {\bibinfo {volume} {77}},\ \bibinfo {pages} {259} (\bibinfo {year}
  {2005})}\BibitemShut {NoStop}%
\bibitem [{\citenamefont {Ohtsuki}\ and\ \citenamefont
  {Ohtsuki}(2016)}]{tomoki2016}%
  \BibitemOpen
  \bibfield  {author} {\bibinfo {author} {\bibfnamefont {T.}~\bibnamefont
  {Ohtsuki}}\ and\ \bibinfo {author} {\bibfnamefont {T.}~\bibnamefont
  {Ohtsuki}},\ }\bibfield  {title} {\bibinfo {title} {Deep learning the quantum
  phase transitions in random two-dimensional electron systems},\ }\href
  {https://10.7566/JPSJ.85.123706} {\bibfield  {journal} {\bibinfo  {journal}
  {Journal of the Physical Society of Japan}\ }\textbf {\bibinfo {volume}
  {85}},\ \bibinfo {pages} {123706} (\bibinfo {year} {2016})}\BibitemShut
  {NoStop}%
\bibitem [{\citenamefont {Carrasquilla}\ and\ \citenamefont
  {Melko}(2017)}]{Carrasquilla2017}%
  \BibitemOpen
  \bibfield  {author} {\bibinfo {author} {\bibfnamefont {J.}~\bibnamefont
  {Carrasquilla}}\ and\ \bibinfo {author} {\bibfnamefont {R.~G.}\ \bibnamefont
  {Melko}},\ }\bibfield  {title} {\bibinfo {title} {{Machine learning phases of
  matter}},\ }\href {https://10.1038/nphys4035} {\bibfield  {journal} {\bibinfo
   {journal} {Nature Physics 2017 13:5}\ }\textbf {\bibinfo {volume} {13}},\
  \bibinfo {pages} {431} (\bibinfo {year} {2017})}\BibitemShut {NoStop}%
\bibitem [{\citenamefont {{Van Nieuwenburg}}\ \emph {et~al.}(2017)\citenamefont
  {{Van Nieuwenburg}}, \citenamefont {Liu},\ and\ \citenamefont
  {Huber}}]{VanNieuwenburg2017}%
  \BibitemOpen
  \bibfield  {author} {\bibinfo {author} {\bibfnamefont {E.~P.}\ \bibnamefont
  {{Van Nieuwenburg}}}, \bibinfo {author} {\bibfnamefont {Y.~H.}\ \bibnamefont
  {Liu}},\ and\ \bibinfo {author} {\bibfnamefont {S.~D.}\ \bibnamefont
  {Huber}},\ }\bibfield  {title} {\bibinfo {title} {{Learning phase transitions
  by confusion}},\ }\href {https://10.1038/NPHYS4037} {\bibfield  {journal}
  {\bibinfo  {journal} {Nature Physics}\ }\textbf {\bibinfo {volume} {13}},\
  \bibinfo {pages} {435} (\bibinfo {year} {2017})}\BibitemShut {NoStop}%
\bibitem [{\citenamefont {Zhang}\ and\ \citenamefont {Kim}(2017)}]{Zhang2017}%
  \BibitemOpen
  \bibfield  {author} {\bibinfo {author} {\bibfnamefont {Y.}~\bibnamefont
  {Zhang}}\ and\ \bibinfo {author} {\bibfnamefont {E.-A.}\ \bibnamefont
  {Kim}},\ }\bibfield  {title} {\bibinfo {title} {Quantum loop topography for
  machine learning},\ }\href {https://10.1103/PhysRevLett.118.216401}
  {\bibfield  {journal} {\bibinfo  {journal} {Phys. Rev. Lett.}\ }\textbf
  {\bibinfo {volume} {118}},\ \bibinfo {pages} {216401} (\bibinfo {year}
  {2017})}\BibitemShut {NoStop}%
\bibitem [{\citenamefont {Ohtsuki}\ and\ \citenamefont
  {Ohtsuki}(2017)}]{Tomi2017}%
  \BibitemOpen
  \bibfield  {author} {\bibinfo {author} {\bibfnamefont {T.}~\bibnamefont
  {Ohtsuki}}\ and\ \bibinfo {author} {\bibfnamefont {T.}~\bibnamefont
  {Ohtsuki}},\ }\bibfield  {title} {\bibinfo {title} {Deep learning the quantum
  phase transitions in random electron systems: Applications to three
  dimensions},\ }\href {https://doi.org/10.7566/JPSJ.86.044708} {\bibfield
  {journal} {\bibinfo  {journal} {Journal of the Physical Society of Japan}\
  }\textbf {\bibinfo {volume} {86}},\ \bibinfo {pages} {044708} (\bibinfo
  {year} {2017})}\BibitemShut {NoStop}%
\bibitem [{\citenamefont {Broecker}\ \emph
  {et~al.}(2017{\natexlab{a}})\citenamefont {Broecker}, \citenamefont
  {Carrasquilla}, \citenamefont {Melko},\ and\ \citenamefont
  {Trebst}}]{Broecker2017}%
  \BibitemOpen
  \bibfield  {author} {\bibinfo {author} {\bibfnamefont {P.}~\bibnamefont
  {Broecker}}, \bibinfo {author} {\bibfnamefont {J.}~\bibnamefont
  {Carrasquilla}}, \bibinfo {author} {\bibfnamefont {R.~G.}\ \bibnamefont
  {Melko}},\ and\ \bibinfo {author} {\bibfnamefont {S.}~\bibnamefont
  {Trebst}},\ }\bibfield  {title} {\bibinfo {title} {{Machine learning quantum
  phases of matter beyond the fermion sign problem}},\ }\href
  {https://10.1038/s41598-017-09098-0} {\bibfield  {journal} {\bibinfo
  {journal} {Scientific Reports 2017 7:1}\ }\textbf {\bibinfo {volume} {7}},\
  \bibinfo {pages} {1} (\bibinfo {year} {2017}{\natexlab{a}})}\BibitemShut
  {NoStop}%
\bibitem [{\citenamefont {Tanaka}\ and\ \citenamefont
  {Tomiya}(2017)}]{Tanaka2017}%
  \BibitemOpen
  \bibfield  {author} {\bibinfo {author} {\bibfnamefont {A.}~\bibnamefont
  {Tanaka}}\ and\ \bibinfo {author} {\bibfnamefont {A.}~\bibnamefont
  {Tomiya}},\ }\bibfield  {title} {\bibinfo {title} {{Detection of phase
  transition via convolutional neural networks}},\ }\href
  {https://10.7566/JPSJ.86.063001} {\bibfield  {journal} {\bibinfo  {journal}
  {Journal of the Physical Society of Japan}\ }\textbf {\bibinfo {volume} {86}}
  (\bibinfo {year} {2017})}\BibitemShut {NoStop}%
\bibitem [{\citenamefont {Broecker}\ \emph
  {et~al.}(2017{\natexlab{b}})\citenamefont {Broecker}, \citenamefont
  {Assaad},\ and\ \citenamefont {Trebst}}]{Broecker2017-1}%
  \BibitemOpen
  \bibfield  {author} {\bibinfo {author} {\bibfnamefont {P.}~\bibnamefont
  {Broecker}}, \bibinfo {author} {\bibfnamefont {F.~F.}\ \bibnamefont
  {Assaad}},\ and\ \bibinfo {author} {\bibfnamefont {S.}~\bibnamefont
  {Trebst}},\ }\bibfield  {title} {\bibinfo {title} {Quantum phase recognition
  via unsupervised machine learning},\ }\href {http://arxiv.org/abs/1707.00663}
  {\bibfield  {journal} {\bibinfo  {journal} {arXiv:1707.00663}\ } (\bibinfo
  {year} {2017}{\natexlab{b}})}\BibitemShut {NoStop}%
\bibitem [{\citenamefont {Ch'Ng}\ \emph {et~al.}(2018)\citenamefont {Ch'Ng},
  \citenamefont {Vazquez},\ and\ \citenamefont {Khatami}}]{Kelvin2018}%
  \BibitemOpen
  \bibfield  {author} {\bibinfo {author} {\bibfnamefont {K.}~\bibnamefont
  {Ch'Ng}}, \bibinfo {author} {\bibfnamefont {N.}~\bibnamefont {Vazquez}},\
  and\ \bibinfo {author} {\bibfnamefont {E.}~\bibnamefont {Khatami}},\
  }\bibfield  {title} {\bibinfo {title} {Unsupervised machine learning account
  of magnetic transitions in the hubbard model},\ }\href
  {https://10.1103/PhysRevE.97.013306} {\bibfield  {journal} {\bibinfo
  {journal} {Phys. Rev. E}\ }\textbf {\bibinfo {volume} {97}} (\bibinfo {year}
  {2018})}\BibitemShut {NoStop}%
\bibitem [{\citenamefont {Zhang}\ \emph {et~al.}(2018)\citenamefont {Zhang},
  \citenamefont {Shen},\ and\ \citenamefont {Zhai}}]{Zhang2018}%
  \BibitemOpen
  \bibfield  {author} {\bibinfo {author} {\bibfnamefont {P.}~\bibnamefont
  {Zhang}}, \bibinfo {author} {\bibfnamefont {H.}~\bibnamefont {Shen}},\ and\
  \bibinfo {author} {\bibfnamefont {H.}~\bibnamefont {Zhai}},\ }\bibfield
  {title} {\bibinfo {title} {Machine learning topological invariants with
  neural networks},\ }\href {https://10.1103/PhysRevLett.120.066401} {\bibfield
   {journal} {\bibinfo  {journal} {Phys. Rev. Lett.}\ }\textbf {\bibinfo
  {volume} {120}} (\bibinfo {year} {2018})}\BibitemShut {NoStop}%
\bibitem [{\citenamefont {Mano}\ and\ \citenamefont
  {Ohtsuki}(2017)}]{Mano2017}%
  \BibitemOpen
  \bibfield  {author} {\bibinfo {author} {\bibfnamefont {T.}~\bibnamefont
  {Mano}}\ and\ \bibinfo {author} {\bibfnamefont {T.}~\bibnamefont {Ohtsuki}},\
  }\bibfield  {title} {\bibinfo {title} {Phase diagrams of three-dimensional
  anderson and quantum percolation models using deep three-dimensional
  convolutional neural network},\ }\href {https://10.7566/JPSJ.86.113704}
  {\bibfield  {journal} {\bibinfo  {journal} {Journal of the Physical Society
  of Japan}\ }\textbf {\bibinfo {volume} {86}} (\bibinfo {year}
  {2017})}\BibitemShut {NoStop}%
\bibitem [{\citenamefont {Carleo}\ and\ \citenamefont
  {Troyer}(2017)}]{Carleo2017}%
  \BibitemOpen
  \bibfield  {author} {\bibinfo {author} {\bibfnamefont {G.}~\bibnamefont
  {Carleo}}\ and\ \bibinfo {author} {\bibfnamefont {M.}~\bibnamefont
  {Troyer}},\ }\bibfield  {title} {\bibinfo {title} {Solving the quantum
  many-body problem with artificial neural networks},\ }\href
  {https://10.1126/SCIENCE.AAG2302} {\bibfield  {journal} {\bibinfo  {journal}
  {Science}\ }\textbf {\bibinfo {volume} {355}},\ \bibinfo {pages} {602}
  (\bibinfo {year} {2017})}\BibitemShut {NoStop}%
\bibitem [{\citenamefont {Sandvik}(1999)}]{Sandvik1999}%
  \BibitemOpen
  \bibfield  {author} {\bibinfo {author} {\bibfnamefont {A.~W.}\ \bibnamefont
  {Sandvik}},\ }\bibfield  {title} {\bibinfo {title} {Stochastic series
  expansion method with operator-loop update},\ }\href
  {https://10.1103/PhysRevB.59.R14157} {\bibfield  {journal} {\bibinfo
  {journal} {Phys. Rev. B}\ }\textbf {\bibinfo {volume} {59}},\ \bibinfo
  {pages} {R14157} (\bibinfo {year} {1999})}\BibitemShut {NoStop}%
\bibitem [{\citenamefont {Orús}(2019)}]{Orus2019}%
  \BibitemOpen
  \bibfield  {author} {\bibinfo {author} {\bibfnamefont {R.}~\bibnamefont
  {Orús}},\ }\bibfield  {title} {\bibinfo {title} {Tensor networks for complex
  quantum systems},\ }\href {https://10.1038/s42254-019-0086-7} {\bibfield
  {journal} {\bibinfo  {journal} {Nature Reviews Physics 2019 1:9}\ }\textbf
  {\bibinfo {volume} {1}},\ \bibinfo {pages} {538} (\bibinfo {year}
  {2019})}\BibitemShut {NoStop}%
\bibitem [{\citenamefont {Hangleiter}\ \emph {et~al.}(2020)\citenamefont
  {Hangleiter}, \citenamefont {Roth}, \citenamefont {Nagaj},\ and\
  \citenamefont {Eisert}}]{Hangleiter2020}%
  \BibitemOpen
  \bibfield  {author} {\bibinfo {author} {\bibfnamefont {D.}~\bibnamefont
  {Hangleiter}}, \bibinfo {author} {\bibfnamefont {I.}~\bibnamefont {Roth}},
  \bibinfo {author} {\bibfnamefont {D.}~\bibnamefont {Nagaj}},\ and\ \bibinfo
  {author} {\bibfnamefont {J.}~\bibnamefont {Eisert}},\ }\bibfield  {title}
  {\bibinfo {title} {Easing the monte carlo sign problem},\ }\href
  {https://10.1126/sciadv.abb8341} {\bibfield  {journal} {\bibinfo  {journal}
  {Science Advances}\ }\textbf {\bibinfo {volume} {6}} (\bibinfo {year}
  {2020})}\BibitemShut {NoStop}%
\end{thebibliography}

\end{document}